# Resting-State EEG Network Profiles Associated with Creative Engagement and Creative Self-Efficacy

Samir Damji[1], Simrut Kurry[2], Shazia'Ayn Babul[3], Joydeep Bhattacharya[4], Naznin Virji-Babul[5]*

[1] Graduate Program in Neuroscience, University of British Columbia, Vancouver, BC, Canada; samirdamji@gmail.com
[2] Department of Occupational Science and Occupational Therapy, University of British Columbia, Vancouver, BC, Canada; simrut.kurry@gmail.com
[3] Mathematical Institute, University of Oxford, Oxford, United Kingdom; shaziaayn@me.com
[4] Academy of Music, School of Creative Arts, Hong Kong Baptist University, Hong Kong SAR, China; jbhattacharya@hkbu.edu.hk
[5] Department of Physical Therapy, Djavad Mowafaghian Centre for Brain Health, University of British Columbia, Vancouver, BC, Canada; naznin.virji-babul@ubc.ca

**Corresponding Author**

*Correspondence: Dr. Naznin Virji-Babul, Department of Physical Therapy, Djavad Mowafaghian Centre for Brain Health, University of British Columbia, 2215 Wesbrook Mall, Vancouver, BC V6T 1Z3, Canada. Email: naznin.virji-babul@ubc.ca

**Short Running Title**

Global EEG Connectivity and Creativity

**Keywords**

resting-state EEG, creativity, graph theory, machine learning, functional connectivity, alpha-band

**Acknowledgements**

This research was supported by a Discovery Grant from the Natural Sciences and Engineering Research Council of Canada (NSERC) to NVB and by a Canada Graduate Scholarships–Master's award from the Canadian Institutes of Health Research (CIHR) to SD.

**Additional Information**

The authors declare no competing interests.

**Abstract**

Creativity is a core cognitive capacity underlying innovation and adaptive problem solving, yet how it is represented in the brain's intrinsic functional architecture is not fully understood. While resting-state fMRI studies have identified large-scale network correlates associated with differences in creativity, EEG provides the temporal resolution for examining oscillatory dynamics contributing to intrinsic network organization. We examined whether resting-state EEG connectivity patterns are associated with individual differences across multiple creativity-related measures. Thirty healthy young adults completed a multidimensional creativity battery comprising the Inventory of Creative Activities and Achievements (ICAA), the Divergent Association Task (DAT), the Matchstick Arithmetic Puzzles Task (MAPT) and a Self-rating (SR) of creative ability. Graph-theoretical analyses of alpha-band functional connectivity revealed two participant groups, each with distinct patterns of neural activity: Cluster 1 showed reduced global connectivity with relatively preserved left frontal connectivity and greater network modularity; Cluster 0 exhibited stronger overall connectivity strength, reduced modularity and higher local clustering. Notably, Cluster 1 reported higher self-rated creative ability and more frequent engagement in real-world creative activities. These findings suggest that resting-state EEG connectivity patterns are associated with variation in creative self-efficacy and creative engagement, highlighting characteristic patterns of alpha-band network organization observed at rest.

## 1. INTRODUCTION

Assessing a person's cognitive traits objectively and quantitatively based on functional brain activity patterns has long been a central aspiration of cognitive neuroscience—one that increasingly demands methods that are not only accurate, but also scalable, fast, and accessible for widespread application [1–3]. Among these traits, creativity, the ability to generate ideas or products that are both novel and useful [4], is of exceptional importance for innovation [5], adaptability [6], and brain health [7]. As artificial intelligence and automation increasingly reshape work [8] and education [9], creativity stands as one of the few uniquely human capacities that drive progress and resilience—a point underscored in the *World Economic Forum's Future of Jobs Report 2025*. Yet, despite decades of behavioural research, the resting-state neural architecture associated with individual differences in creativity is not well understood.

There is accumulating evidence that the brain's intrinsic functional architecture, reflected in spontaneous, ongoing activity in the absence of external tasks, provides insights into relatively stable differences in cognitive functioning [10,11]. The brain at rest is not at rest: resting-state (RS) functional interactions, which emerge from coherent patterns of neural activity during wakeful rest, are thought to reflect the brain's baseline modes of information integration and internal mentation and serve as the basis for task-based neural activation [12,13]. With advances in machine learning and network neuroscience, researchers have increasingly been able to model and decode these intrinsic connectivity patterns to identify patterns associated with  intelligence [14], personality [15], and clinical characteristics [16,17].

Parallel advances in resting-state fMRI have begun to reveal how the brain's intrinsic network architecture relates to creative ability.  Studies using graph-theoretical and connectivity approaches have shown that higher levels of creativity are associated with more flexible modular organisation of the default-mode network, reduced coupling of temporal regions, and enhanced cross-network communication patterns supporting integration and segregation across cognitive systems [18]. In line with this view, a creativity-related resting-state network was identified to have higher network efficiency in highly creative individuals, suggesting more efficient intrinsic information transfer at rest [19]. More recent work indicates that creative ability is predicted by resting-state functional connectivity properties of control–default connector hubs and by dynamic switching between the default-mode and executive control networks at rest, highlighting flexible cross-network coordination as a core feature of creative cognition [20,21]. Other work focusing on subcortical–cortical interactions has revealed that variations in dopaminergic midbrain and striatal connectivity at rest are associated with creativity-related traits linked to openness and ideational fluency, suggesting that neuromodulatory systems contribute to spontaneous associative processing [22]. Investigations of cerebral blood flow and network physiology further indicate that reduced baseline metabolic activity in the precuneus, a core default-mode hub, accompanies higher creativity, consistent with interpretations that lower resting activation within introspective networks may facilitate flexible reconfiguration during thought generation [23]. More recent evidence from creative experts has shown attenuated connectivity between visual and higher-order association regions, interpreted as a state of perceptual decoupling that favours internal ideation [24]. Relatedly, individual differences in creativity have also been linked to the unimodal–transmodal functional connectivity gradient, with higher trait creativity associated with increased separation along the sensory-to-associative axis of intrinsic functional organization at rest [25]. Collectively, these studies point to distinctive resting-state configurations, particularly within the default-mode, frontoparietal control, and visual networks, as neural correlates of creative potential.

While fMRI research has advanced our understanding of the large-scale network architecture underlying creative ability, electroencephalography (EEG) offers a complementary, yet scalable, approach with millisecond-level temporal precision, allowing investigation of the fast oscillatory dynamics that shape intrinsic brain organisation at rest. Early EEG studies provided some of the first evidence that creativity is reflected in the brain's resting functional activity. Analysis of spectral power and interregional coherence revealed that individuals with higher creative potential exhibited reduced inter-areal coupling, particularly across right-hemisphere networks during eyes-open rest, suggesting a state of relative neural decoupling associated with flexible associative processing [25]. Two decades later, this idea was extended by examining the temporal dynamics of alpha-band amplitude fluctuations, showing that highly creative individuals displayed lower long-range temporal correlations over right central and temporal scalp regions, indicative of more rapid and adaptive state switching within intrinsic oscillatory networks [26]. Subsequent work using EEG microstate analysis further linked creativity to distinct patterns of resting-state topographical stability, where creativity-related differences in the duration and transition probabilities of visual and control microstates highlighted the importance of spontaneous network transitions as markers associated with creative traits [27]. Most recently, high-density EEG combined with connectome-based predictive modelling identified reproducible gamma-band connectivity patterns that successfully predicted individual creativity scores both within-sample and in an independent validation cohort, demonstrating the predictive value of resting-state EEG for creative potential [28].

Despite these promising advances, the current literature on resting-state EEG and creativity remains conceptually narrow and methodologically fragmented. Most studies have adopted single-task or single-dimension approaches, typically using divergent thinking scores as the sole indicator of creativity, which constrains interpretation of neural findings to a limited facet of creative ability rather than the broader construct encompassing ideation, problem solving, and everyday creative engagement. In addition, previous studies have largely focused on isolated EEG metrics such as power, temporal autocorrelation, or microstate duration, each capturing only one aspect of intrinsic brain dynamics without integrating them within a unified network framework. As a result, it remains unclear whether the brain's global functional organisation at rest, reflected in its overall pattern of connectivity and topological structure, can meaningfully differentiate individuals in terms of creativity-related characteristics. Furthermore, existing studies have tended to rely on group-level correlational analyses, which assume linear associations between neural indices and creativity scores, thereby overlooking the possibility that distinct neurocognitive profiles might underlie similar levels of creative performance. Addressing these conceptual gaps requires a data-driven approach that can identify naturally emerging subgroups based on whole-brain resting-state connectivity and relate these to multiple dimensions of creativity, encompassing self-efficacy, behavioural performance, and real-world engagement.

The present study was designed to address these conceptual gaps by examining whether individual differences in creativity are reflected in the global organisation of intrinsic brain networks as measured by resting-state EEG. Moving beyond prior correlational approaches, we adopted a data-driven network perspective, using measures of large-scale functional connectivity to identify participant subgroups characterised by distinct intrinsic brain architectures. We then examined how these neural subgroups differed across a comprehensive battery of creativity measures, encompassing divergent and convergent thinking, real-world creative engagement, and self-perceived creative ability. This multidimensional framework allowed us to test whether the brain's spontaneous activity at rest captures patterns associated with both self-perceived creativity, creative engagement and creative behaviour.

We hypothesised that distinct participant groups would emerge based on the topology of their resting-state EEG networks, reflecting qualitatively different modes of intrinsic brain organisation. We further expected that one group would demonstrate higher scores on selected creativity-related measures. Finally, we predicted that this subgroup would exhibit reduced global connectivity strength but greater modular segregation in the alpha-band (8-13 Hz), a pattern previously associated with flexible network organization. This prediction is grounded in both methodological and theoretical considerations: alpha-band activity is known to exhibit high signal quality and reliability across individuals during eyes-closed resting-state EEG recordings relative to other frequency bands [29,30], making it well-suited for network structure analysis. Additionally, a broad literature has linked alpha-band oscillations to internally directed attention and creative cognition [31,32], further motivating its inclusion as a key frequency band of interest in the present study.

Together, these aims situate the present study within a broader effort to understand how the brain's spontaneous activity at rest is associated with individual differences in cognition, providing a framework for linking intrinsic network organization to creativity-related characteristics.

## 2. MATERIALS AND METHODS

### 2.1. Participants

Thirty undergraduate students (mean age = 21.17, SD = 1.16; 15 women) between the ages of 19-25 years were recruited from the University of British Columbia (UBC). Inclusion criteria included fluency in English, being right-handed, and free of any neurological conditions that could affect brain function (e.g., stroke, concussion, epilepsy, or neurodevelopmental disorders). Four participants were excluded due to technical difficulties during EEG recording, resulting in a final analyzed sample of 26 healthy young adults (mean age = 21.13, SD = 1.15; 13 women). The study was approved by the University of British Columbia Behavioural Research Ethics Board (H22-01386) in accordance with the Helsinki declaration. All research was performed in accordance with relevant guidelines/regulations. All participants provided informed consent as per the guidelines of the Human Ethics Review Board of the University of British Columbia. All received financial compensation for their travel and parking.

### 2.2. Behavioral Creativity Tasks

Participants first provided a rating of self-perceived creative ability to index self-efficacy. They were then asked to complete three standardized creative thinking assessments administered in a fixed order. These standardized assessments captured complementary facets of creative cognition—real-world creative engagement, verbal divergent thinking, and visuospatial convergent thinking—alongside self-perceived creative ability. The creative assessments included:

1. The Inventory of Creative Activities and Achievements (ICAA) [33] captures real-world creative engagement across eight domains (literature, music, arts-and-crafts, creative cooking, sports, visual arts, performing arts, science, and engineering) over the past ten years. It provides three scores: Activities, Achievements, and a Total score (sum of both). The Activities subscale measures the frequency of participation in creative activities, and the Achievements subscale measures the level of attained recognition within each domain. Accordingly, the Total score reflects overall engagement in everyday creative pursuits, with higher scores reflecting more frequent and deeper involvement in creative activities.
2. The Divergent Association Task (DAT) [34] offers a measure of verbal divergent thinking. Participants are asked to generate ten words that are as different from each other as possible in both meaning and usage. The task performance is measured by the mean pairwise semantic distance between words using a large-scale word-embedding model derived from natural-language corpora (word2vec-like architecture). Higher scores indicate greater conceptual diversity and flexibility.
3. The Matchstick Arithmetic Puzzles Task (MAPT) [35,36] provides a measure of visuospatial convergent thinking. Participants are presented with a series of incorrect arithmetic equations composed of Roman numerals and operators from matchsticks (e.g., XI = X + I). Each equation can be made mathematically correct by moving a single matchstick. Type A and B problems typically require minimal representational change, whereas Type C and D problems demand deeper restructuring of learned rules. The task performance is measured by the solution rate per problem type.

4. Self-rating (SR) of creative ability: Participants rated their own creative ability on a single Likert item: *"Creativity is the ability to generate novel and useful ideas. On a scale of 1 to 7, how creative would you subjectively say you are, based on this definition?"* Participants responded on a 7-point scale, where 1 = "not creative at all," 4 = "moderately creative," and 7 = "extremely creative."

Collectively, these four measures provide a broad index of both objective creative performance (DAT, MAPT) and subjective creative engagement (ICAA, SR).

## 2.3. EEG Acquisition and Preprocessing

Following completion of the behavioral assessments, five minutes of resting-state EEG were recorded. All data was collected at the Perception-Action Lab at UBC. Participants sat comfortably with their eyes closed and were instructed to relax, remain still, and avoid structured thought. EEG data were collected using a 64-channel HydroCel Geodesic Sensor Net (HCGSN 64; Electrical Geodesics, Inc., Eugene, OR) connected to a Net Amps 400 amplifier at a sampling rate of 500 Hz with Cz as the online reference. All electrode impedances were kept below 50 kΩ.

Raw EEG data were preprocessed using EEGLAB (v2023.1) [37] in MATLAB. Each participant's EEG was re-referenced to the average of all channels and down sampled to 250 Hz. A notch filter at 60 Hz, a low-pass filter at 50 Hz, and a high-pass filter at 0.5 Hz were applied to the EEG. Independent Component Analysis (ICA) was used to identify and remove any non-brain artifacts through a combination of examining the topographic, power, and time series properties of components and EEGLAB's ICLabel classification function [38]. On average, 21 ± 3 components were removed per participant using the *runica* algorithm (an implementation of the Infomax ICA method) [37], and all datasets were subjected to a final visual inspection to confirm the removal of residual artifacts.

## 2.4. Functional Connectivity Analysis

Functional connectivity was assessed using the debiased weighted Phase Lag Index (wPLI), which quantifies non-zero phase lag coupling between signals and minimises the influence of volume conduction and common sources [39]. wPLI ranges from 0 (no consistent phase-lead/lag) to 1 (fully consistent phase delay). As the analysis was restricted to sensor space, wPLI was preferred as an appropriate functional connectivity metric because, unlike zero-phase-lag approaches such as magnitude-squared-coherence or phase synchrony that are inflated by instantaneous signal spread [40], wPLI quantifies the imagery component of the cross-spectrum, therefore capturing only non-zero-lag phase relationship and thus providing a conservative estimate of genuine inter-regional communication.

For each participant, we computed cross-spectral density matrices using multitaper spectral decomposition with a discrete prolate spheroidal sequence tapers (time-bandwidth product = 6; K = 11 tapers) and ±1 Hz spectral smoothing. Connectivity estimates were computed across canonical frequency bands: delta (1–4 Hz), theta (4–8 Hz), alpha (8–13 Hz), beta (13–30 Hz), and gamma (30–45 Hz). For completeness, analyses were also conducted in sub-bands to examine potential frequency-specific sensitivity, motivated by prior work suggesting that lower- (8–10 Hz) and upper-alpha (10–13 Hz) sub-bands may differentially represent creativity-related cognitive processes [26,41,42]. For each epoch (6 sec long), wPLI values were computed between all pairs of the 64 electrodes and then averaged across epochs to yield a single 64 × 64 connectivity matrix per participant.

Connectivity matrices were inspected visually and averaged across participants to generate group-level heatmaps. The upper triangular values (excluding the diagonal) were extracted and vectorized (2016 unique edges) for subsequent analysis.

### 2.5. Graph Similarity Analysis

To quantify similarity in global network organisation across participants, each 64x64 wPLI matrix was treated as a weighted, undirected graph with nodes corresponding to electrodes and edges representing connectivity strengths. For graph analysis, we employed the NetworkX Python library [43]. Between-participant graph similarity was evaluated with DeltaCon, a principled distance metric that captures both local and global changes by comparing node-to-node affinity matrices generated via the fast belief propagation algorithm [44]. Prior studies have validated DeltaCon for comparing large-scale brain networks [45,46]. Pairwise DeltaCon distances were computed for all subject graphs using the netrd Python library [47], resulting in an $n \times n$ symmetric distance matrix ($n = 26$). Lower DeltaCon distance values indicate greater topological similarity. This approach avoids arbitrary thresholding of individual graphs, preserves edge-weight information, and provides a continuous measure of inter-subject similarity suitable for unsupervised clustering.

### 2.6. Dimensionality Reduction and Clustering

The DeltaCon distance matrix represents a non-Euclidean space describing pairwise structural dissimilarities between participants' functional connectivity networks. To visualise and cluster these relationships, we applied Uniform Manifold Approximation and Projection (UMAP), a nonlinear dimensionality reduction technique that preserves both local and global structure of high-dimensional data (*n_components* = 2, *metric* = 'precomputed', *n_neighbors* = 15, *min_dist* = 0.1, *random_state* = 52) [48]. Following dimensionality reduction, k-means clustering was applied on the 2-D embedding to identify subgroups of participants with similar brain network organization (*n_init* = *"auto", random_state* = *52*) with candidate solutions restricted to $k \leq 5$ due to sample size. The optimal number of clusters was determined using the silhouette method [49], which quantifies how well-defined each cluster is by computing a silhouette score for each data point. Higher average silhouette values indicate both greater within-cluster cohesion and stronger separation between clusters, making this a robust and widely used criterion for evaluating clustering quality. Multiple candidate solutions were evaluated, with consideration given to both cluster quality and downstream interpretability in the context of this preliminary investigation.

Clustering robustness was evaluated using the Adjusted Rand Index (ARI) across 100 random seeds, a conservative, chance-corrected measure of agreement with baseline cluster assignments. ARI values range from −1 to 1, with an ARI value of 0 indicating chance-level agreement across random seeds. Stability to algorithmic randomness was assessed by perturbing random seeds for both the full UMAP–k-means pipeline and k-means clustering on fixed embeddings, while subject-level sensitivity was evaluated using a supplementary leave-one-out (LOO) analysis with random seeds held fixed. Together, these analyses quantified robustness to both stochastic initialization and individual-subject influence.

The full EEG analysis pipeline, detailed in Sections 2.4–2.6, is shown schematically in Figure 1A-E.

**Figure 1.** *Analysis pipeline for clustering resting-state EEG based on graph similarity.*

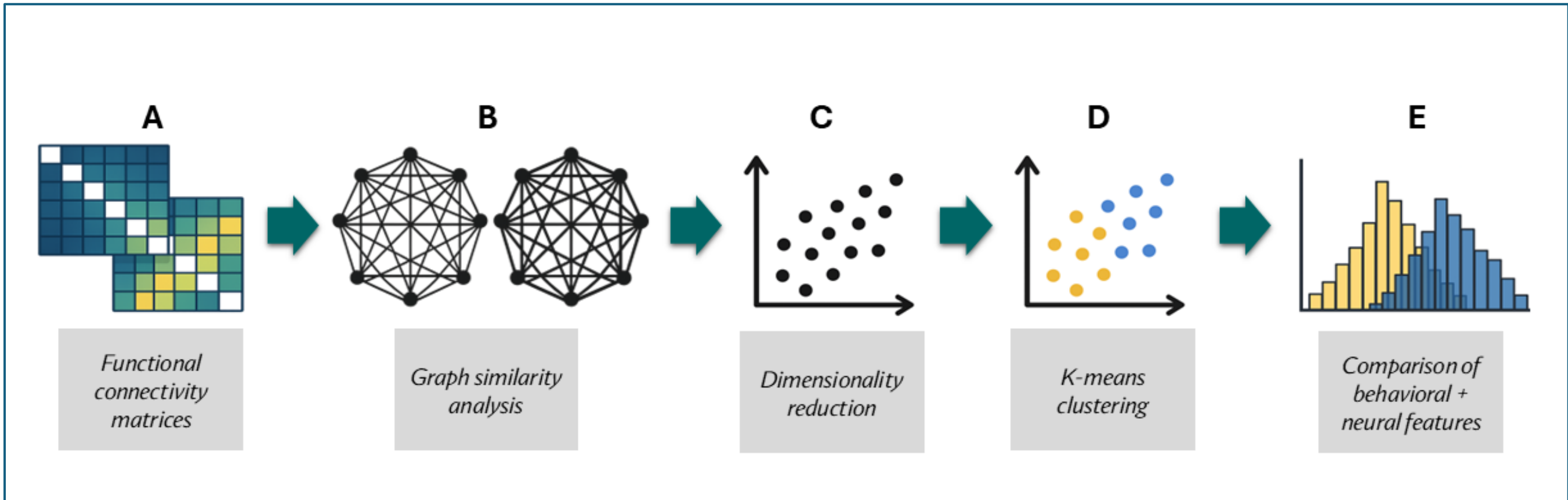

## 2.7. Statistical Analysis

Group differences in behavioral creativity across clusters were identified in a data-driven manner from functional connectivity profiles. Statistical tests were selected according to the optimal number of clusters and the distributional properties of each creativity measure. The ICAA, DAT, and self-reported ability measures yielded continuous numeric scores, whereas the MAPT produced binary outcomes (solved vs. not solved) for each question. For the MAPT, the solution rate was calculated for each of the four problem types (A-D), defined as the proportion of participants who successfully solved at least one problem of that type. Accordingly, we compared MAPT performance across clusters based on solution rates per problem category.

As a secondary analysis of the behavioral data, a series of non-parametric permutation tests were conducted to assess the likelihood that any observed cluster differences in behavioral scores across EEG-defined clusters occurred by chance. The underlying question addressed by this method is: "*Are the observed between-cluster differences in creativity scores larger than what would be expected by random chance if cluster labels were assigned arbitrarily?"* Cluster labels were randomly shuffled 10,000 times—preserving both the original creativity scores and the cluster sizes—to generate an empirical null distribution of the relevant test statistic. This Monte Carlo–based approach does not rely on distributional assumptions (e.g., normality or equal variance) and is well-suited for small sample sizes. Two-tailed empirical *p*-values were computed as the proportion of permutations where the simulated test statistic was greater than or equal to the observed test statistic in absolute value. A two-tailed approach was preferred because cluster labels were data-driven; we tested for any difference in behavioral measures derived from neural data, and thus a non-directional null hypothesis represented the most conservative approach.

To assess how average brain activity patterns corresponded to differences in creative ability, self-perception, and engagement, the average functional connectivity (FC) matrix was computed across clusters and compared at the global, edgewise, and nodal levels. For edgewise comparisons, appropriate multiple comparison correction procedures were applied to statistically control for false positives. At the node level, node strength was computed for each of the 64 electrodes in the group-level FC matrices and compared between clusters. Node strength, defined as the sum of all edge weights connected to a given node, provides a local measure of that node's integration within the broader network. In the context of a weighted, undirected network, node strength serves as an appropriate alternative to node degree, which captures the number of connections in binary graphs (i.e., edge presence or absence is quantified). While DeltaCon quantifies global graph similarity via fast belief propagation, node strength offers a complementary, node-level perspective on network topology.

Finally, a set of weighted global graph-theoretical metrics were computed for each subject's FC matrix, including mean node strength, characteristic path length, global efficiency, mean clustering coefficient, and modularity. Mean node strength quantifies the average total connectivity load per node in a weighted network. As opposed to degree centrality, which reflects the average number of connections per node in a binary (unweighted) graph, mean node strength incorporates the magnitude of those connections, making it a more informative measure for weighted, undirected EEG-derived networks. Characteristic path length and global efficiency are complementary path-based metrics of global integration. Both rely on defining distance as the reciprocal of edge weight (1/weight), ensuring stronger connections are treated as shorter paths. Characteristic path length reflects the arithmetic mean of the shortest path length between all node pairs in the network, whereas global efficiency reflects the harmonic mean of the inverse shortest path lengths. In practice, global efficiency is approximately the inverse of characteristic path length, with higher efficiency corresponding to shorter path lengths. Modularity was calculated using the weighted greedy modularity optimization algorithm [50], which identifies community structure by maximizing within-module connectivity relative to chance. These metrics were selected to further characterize network structure (i.e., integration and segregation), enabling a multiscale characterization of cluster differences in brain network architecture.

# 3. RESULTS

## 3.1. Final Sample Characteristics

After data quality control, 26 participants (13 female) were retained for all analysis. Full demographic characteristics and descriptive statistics of the final sample are presented in Table 1.

**Table 1.** *Overall sample demographics and behavioral profile (N = 26).*

| Measure | Value |
|---|---|
| **N** | 26 |
| **Age (M ± SD)** | 21.1 ± 1.2 |
| **% Male (M/F)** | 50.0% (13M/13F) |
| **Academic Discipline Distribution (%)** | |
| *Natural Sciences* | 38.5 |
| *Engineering/Applied Sciences* | 34.6 |
| *Social Sciences* | 11.5 |
| *Business/Management* | 15.4 |
| **SR of creative ability** | 4.46 ± 0.76 |
| **ICAA: Total** | 169.19 ± 47.31 |
| **ICAA: Achievements** | 92.42 ± 32.80 |
| **ICAA: Activities** | 76.77 ± 24.86 |
| **DAT** | 81.10 ± 6.61 |

| Solved MAPT: Hard problem (%) | 53.8 |
|---|---|
| **Solved MAPT: Type C problem (%)** | 23.1 |
| **Solved MAPT: Type D problem (%)** | 42.3 |

Across the full sample, mean self-perceived creativity scores indicated a moderate level of perceived ability ($M = 4.46 \pm 0.76$ on a 7-point scale). Mean scores for the Inventory of Creative Activities and Achievements (ICAA) and the Divergent Association Task (DAT) were slightly above published normative averages, whereas solution rates on the Matchstick Arithmetic Puzzles Task (MAPT) indicated expected ceiling performance on Type A/B problems, which involve minimal constraint relaxation, and variability emerged on the more difficult Type C/D problems. The latter provided a sensitive measure of individual differences in convergent problem-solving.

## 3.2. Identification of EEG-derived Clusters

We focused on the alpha frequency band (8–13 Hz) because it exhibited the most stable functional connectivity patterns across participants and is consistently linked to internally directed attention and creative cognition [26,31,32,51–53]. The relatively weaker and irregular global connectivity in other frequency bands produced sparser graphs, preventing DeltaCon from reliably computing distance values for all pairwise comparisons across our sample. This supported our hypothesis and guided our choice of the alpha-band for subsequent clustering analyses.

Clustering of resting-state functional connectivity matrices using DeltaCon graph distances revealed two distinct sub-groups (Figure 2). The silhouette coefficient identified $k = 2$ as the optimal partition (Figure 2C). Within-cluster DeltaCon distances were significantly smaller than between-cluster distances (within = 3.48; between = 4.92; $t(235.02) = -8.49$, $p < 0.001$, $d = 0.94$), confirming that the two-cluster solution captured meaningful differences in global network topology. Given that the upper-alpha-band produced the strongest differentiation with behavioral creativity measures, all alpha-band clustering results reported below are based on upper-alpha (10–13 Hz) functional connectivity. Clustering in the full alpha-band yielded qualitatively similar patterns, whereas effects in the lower-alpha band were less pronounced. The resulting clusters, Cluster 0 and Cluster 1, were demographically comparable (Table 2).

Moderate agreement was observed when perturbing the full UMAP–k-means pipeline (mean ARI = 0.62 ± 0.17; median ARI = 0.70), reflecting moderate variability due to stochasticity in nonlinear dimensionality reduction, as expected (Figure 2E). Higher agreement was observed for k-means clustering with fixed UMAP embeddings (mean ARI = 0.73 ± 0.31; median ARI = 0.85). Together, these results indicate that cluster structure is largely preserved across runs, with variability driven by a small number of borderline subjects, consistent with a beyond-chance, robust clustering solution given the small-sample, healthy adult EEG network analysis context. As a supplementary sensitivity analysis, leave-one-out (LOO) perturbation of the full UMAP–k-means pipeline yielded comparable stability (mean ARI = 0.65 ± 0.14; median ARI = 0.69; min/max: 0.44/1.00), indicating that cluster structure was not unduly driven by any single subject.

**Figure 2.** *Graph similarity analysis results.*

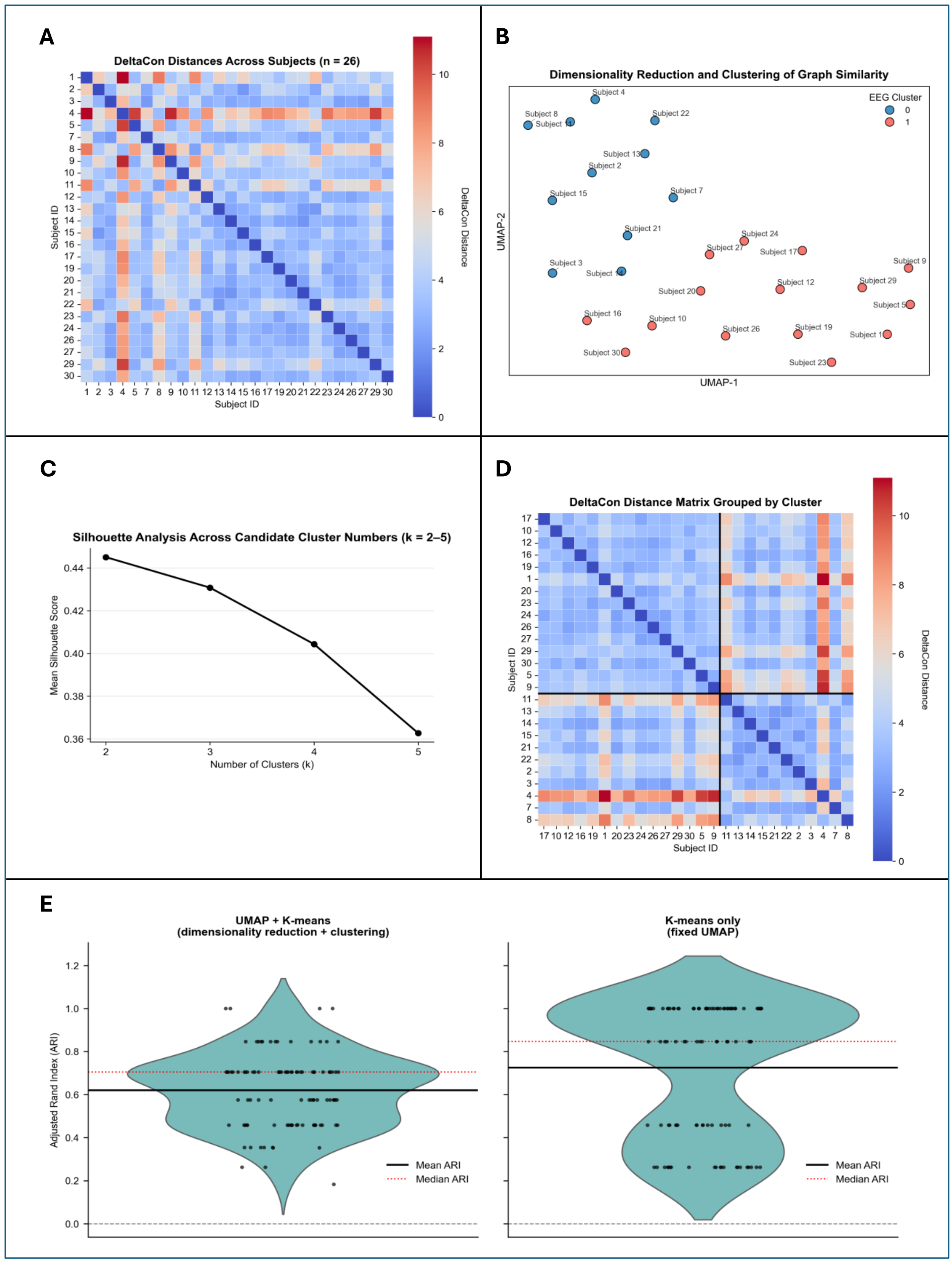

**Table 2.** *Demographic characteristics of the EEG-derived clusters.*

| **Measure** | **EEG Cluster 0** | **EEG Cluster 1** |
|---|---|---|
| **N** | 11 | 15 |
| **Age (M ± SD)** | 20.8 ± 1.0 | 21.4 ± 1.2 |
| **% Male (M/F)** | 45.5 (5M/6F) | 53.3 (8M/7F) |
| **Academic Discipline Distribution (%)** | | |
| *Natural Sciences* | 36.4 | 40.0 |
| *Engineering/Applied Sciences* | 36.4 | 33.3 |
| *Social Sciences* | 18.2 | 6.7 |
| *Business/Management* | 9.1 | 20.0 |

### 3.3. Cluster Level Creativity Profiles

Although the two EEG-derived clusters were demographically similar, they differed in creativity scores (Figure 3: Cluster 1 – red, Cluster 0 – blue).

**Figure 3.** *Behavioral creativity profiles across the two clusters.*

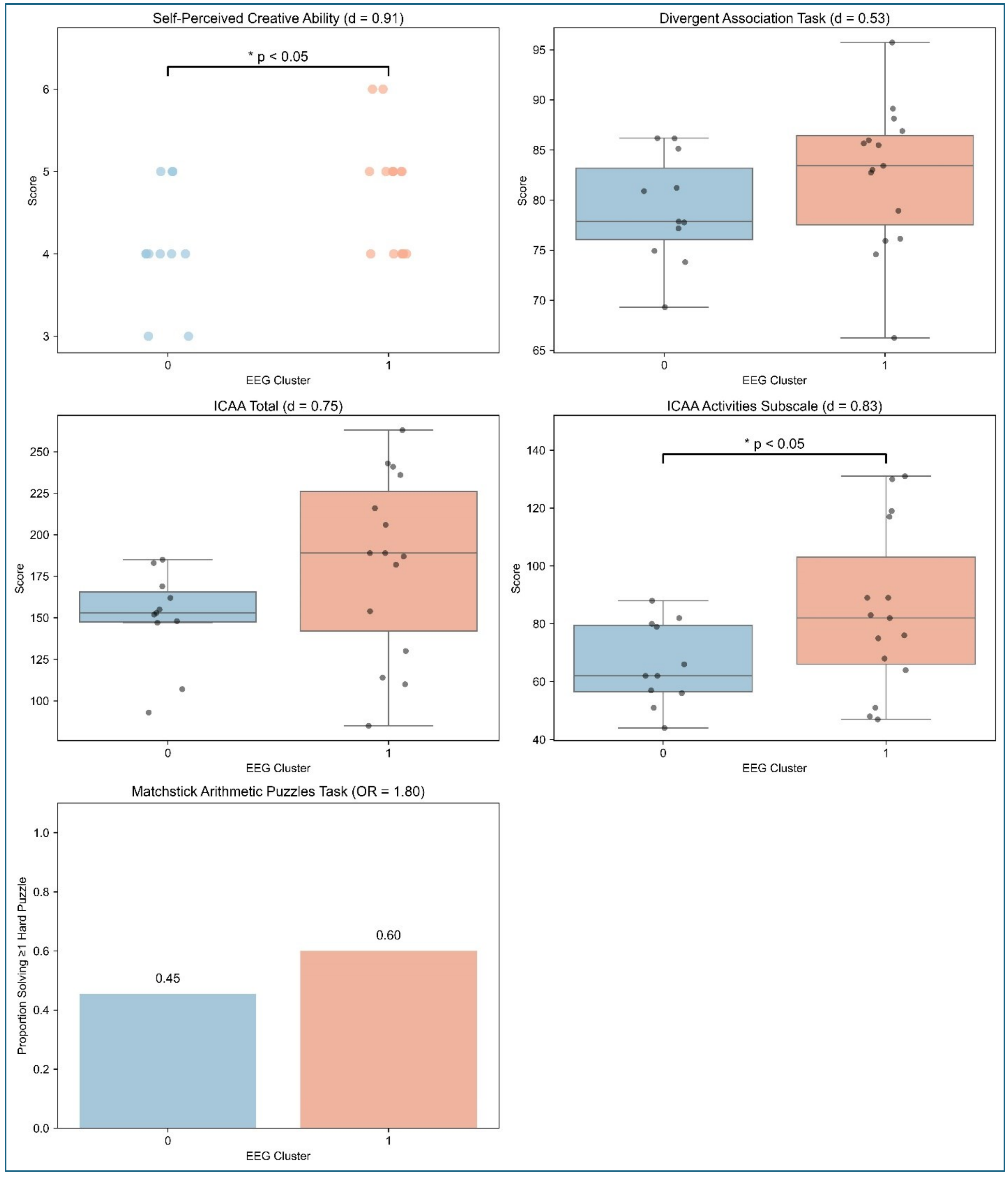

Participants in Cluster 1 reported significantly higher self-perceived creative ability ($M$ = 4.73 ± 0.70) than Cluster 0 ($M$ = 4.09 ± 0.70), $t(21.76) = 2.31$, $p = .03$, $d = 0.92$. The mean difference in SR of creative ability remained significant under permutation testing ($p = .04$).

DAT scores were higher in Cluster 1 ($M$ = 82.54 ± 7.20) than in Cluster 0 ($M$ = 79.13 ± 5.40), however, the differences did not reach significance ($t$(23.97) = 1.38, $p$ = .18, $d$ = 0.535). The corresponding permutation test of the observed mean difference likewise did not reach significance ($p$ = .20).

On the ICAA Activities subscale, Cluster 1 reported engaging in everyday creative activities more frequently ($M$ = 84.60 ± 28.34) than those in Cluster 0 ($M$ = 66.09 ± 14.24), and the difference was statistically significant ($t$(21.70) = 2.18, $p$ = .04, $d$ = 0.825), indicating a large effect size. On the ICAA Achievements subscale, Cluster 1 reported higher levels of publicly recognized creative accomplishments ($M$ = 98.40 ± 37.41) than those in Cluster 0 ($M$ = 84.27 ± 24.55), though the difference did not reach statistical significance ($t$(23.79) = 1.16, $p$ = .26, $d$ = 0.446). Combining the Activities and Achievements subscales, Cluster 1 displayed a higher ICAA Total score ($M$ = 183.00 ± 54.26) than Cluster 0 ($M$ = 150.36 ± 28.23), and the difference was marginally significant ($t$(22.04) = 1.99, $p$ = .06, $d$ = 0.755), indicating a large effect size. In the permutation analysis, mean differences in ICAA Activities ($p$ = .06) and ICAA Total ($p$ = .08) approached significance, whereas the ICAA Achievements subscale ($p$ = .29) showed weaker evidence for group separation.

For MAPT Type D problems, a similar pattern was observed: 53% of Cluster 1 versus 27% of Cluster 0 solved at least one item (OR = 3.05). For Type C problems, the success rates were 27% vs. 18%; OR = 1.64. When both problem types were combined, the overall success rate remained higher in Cluster 1 (60% vs. 46%; OR = 1.80). However, these differences did not reach statistical significance according to two-tailed Fisher's exact tests (all $p > .24$). The permutation test, based on the observed difference in percentage solution rates between clusters, similarly showed no significant effects (all $p > .24$). Nonetheless, Cluster 1 exhibited higher solution rates across all binary MAPT outcomes in our sample (Type C, Type D, Type C and/or D combined).

## 3.4. Neural Connectivity Patterns

### *3.4.1. Global Connectivity Strength*

Mean alpha-band functional connectivity (FC) across 2016 edges differed between clusters (Figure 4). Cluster 1 showed significantly lower global mean FC ($M$ = 0.225) than Cluster 0 ($M$ = 0.391), $t$(15.88) = -3.92, p = .0012, $d$ = 1.66.

**Figure 4.** *Mean functional connectivity difference matrix (Cluster 1 – Cluster 0).*

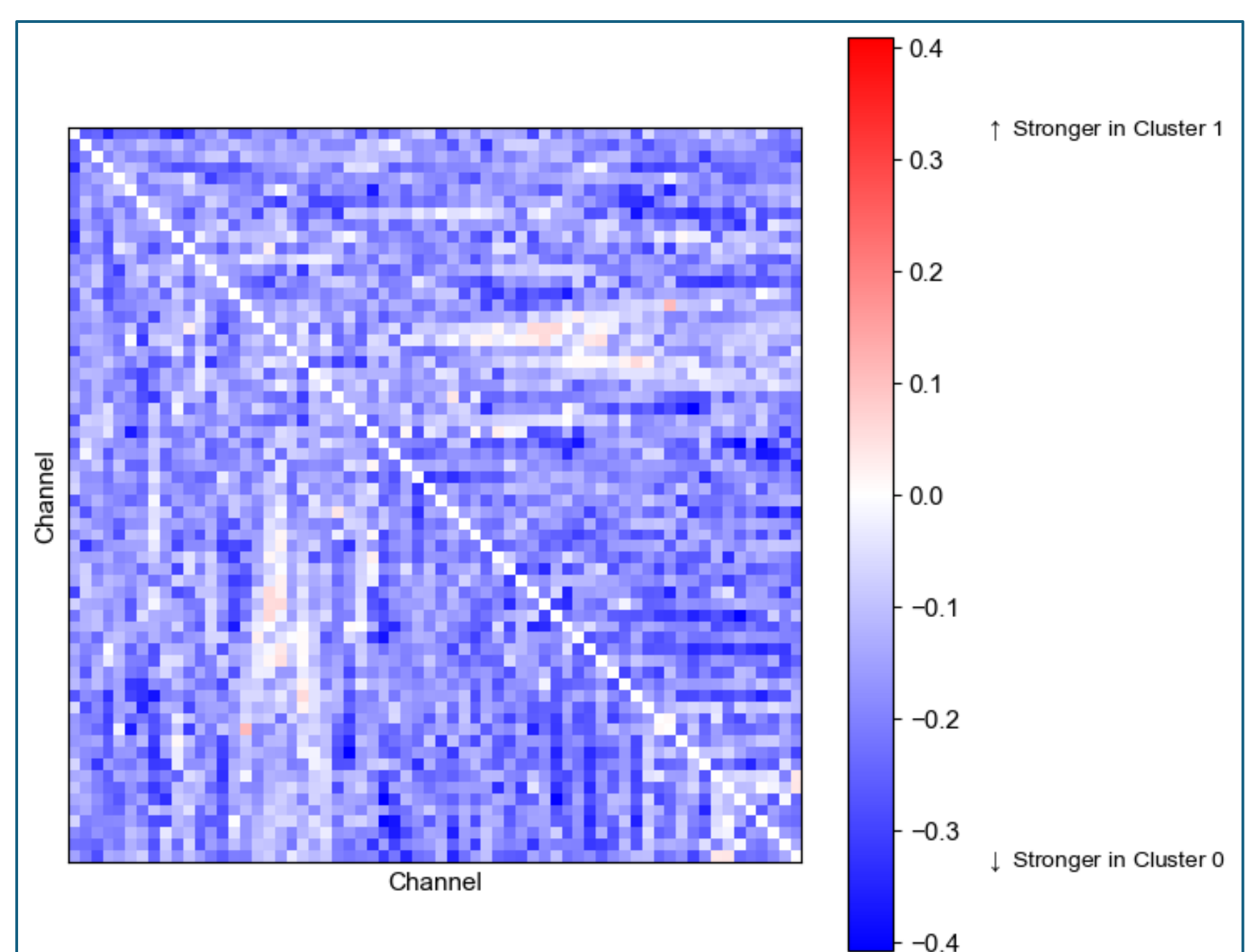

*3.4.2. Edgewise Comparisons*

Edgewise comparisons across 2,016 connections identified 10 significant edges after false discovery rate (FDR) correction ( $\alpha$ = 0.03). All significant edges displayed higher connectivity in Cluster 0. The mean effect size across these edges was *d* = 2.14 (range = 1.83–2.61). No significant edges were stronger in Cluster 1 (Figure 5A).

*3.4.3. Nodal Strength*

Nodal strength maps (sum of edge weights per node) indicated widespread hypoconnectivity in Cluster 1 relative to Cluster 0, with relatively preserved strength over left-frontal electrodes region (Figure 5B).

**Figure 5.** *Topographical edgewise and nodal connectivity differences between clusters.*

A

Significant Edgewise Differences, FDR-Corrected with α = 0.03

B

$$s_i = \sum_{\substack{j=1 \\ j \neq i}}^{N} w_{ij}$$

where $s_i$ = node strength of electrode i, $w_{ij}$ = wPLI between electrodes i and j

*3.4.4. Global Graph Architecture*

At the global level (Figure 6), Cluster 1 exhibited lower node strength ($p$ = .0012, $d$ = 1.61), longer characteristic path length ($p$ = .0153, $d$ = 1.00), reduced global efficiency ($p$ = .0078, *d* = 1.23), greater modularity (*p* < .001, *d* = 1.82) and lower clustering coefficient (*p* < .001, *d* = 2.16). Notably, all these effects were large. Together, these findings indicate that Cluster 1's brain networks are characterized by reduced global synchronization but enhanced modular organization—suggesting a more segregated, flexible, and potentially adaptive network configuration.

**Figure 6.** *Distributions of global graph metrics within EEG clusters.*

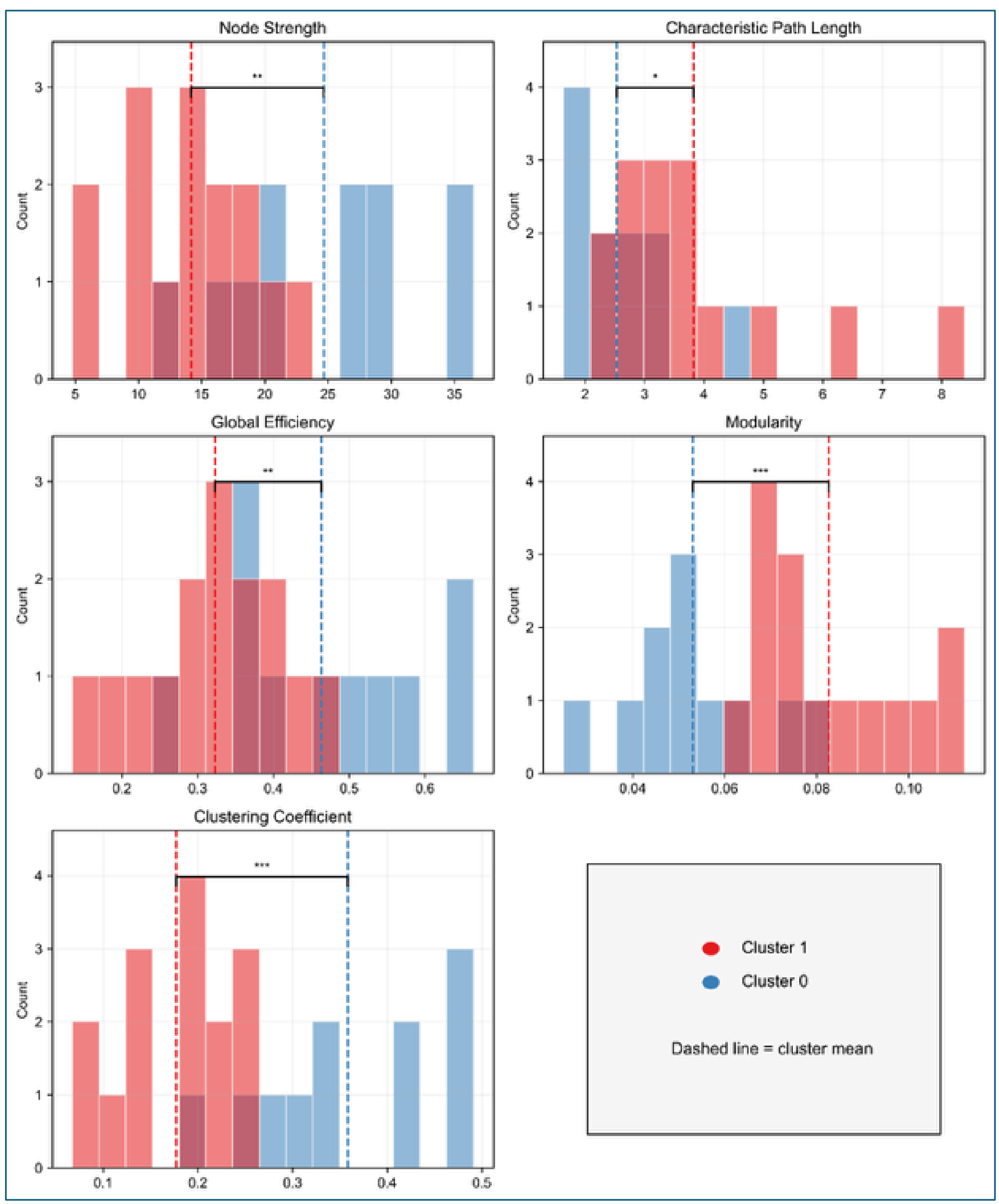

## 4. DISCUSSION

This study tested whether creativity, a multidimensional human capacity involving both spontaneous and controlled cognitive processes [54,55], is associated with features of the brain's resting-state functional organisation. By applying a data-driven, graph-theoretical approach to resting-state EEG, we identified two neurofunctionally distinct participant groups that differed in their levels of self-perceived creativity and creative activity engagement. Consistent with our hypotheses, the higher-scoring group, Cluster 1, exhibited (i) reduced global alpha-band connectivity, (ii) relatively preserved left-frontal coupling, and (iii) greater modular segregation of brain networks relative to Cluster 0. Together, these findings extend recent network-based accounts of creativity by showing that individual differences in creativity-related constructs are associated with the topology of spontaneous brain activity.

### 4.1. Emergent Neural Subgroups Reflecting Network Architecture

Supporting our first hypothesis, unsupervised clustering of whole-brain connectivity revealed two relatively stable neural subgroups despite demographic similarities. While additional substructure cannot be ruled out, this preliminary investigation was designed to model a primary axis of variation in intrinsic network organization related to creativity, rather than finer-grained neurobehavioral phenotypes that would be underpowered in a modest sample. Importantly, cluster robustness analyses indicated generally stable solutions, consistent with overlapping rather than rigidly discrete neural architectures. This pattern underscores that the observed grouping reflects meaningful, but non-deterministic, differences in large-scale network organization, and highlights sources of instability that may be informative for future, larger-scale investigations.

The identification of these EEG clusters demonstrates that intrinsic connectivity patterns alone can capture meaningful cognitive heterogeneity, echoing recent efforts to identify data-driven biotypes in perception, attention and intelligence [56–58]. These findings support the view that  resting-state networks reflect reproducible patterns associated with individual differences in cognition [59–61]. Extending this framework to creative cognition, our results show that the brain's intrinsic organisation varies systematically across individuals prior to task engagement, consistent with recent fMRI evidence linking creativity-related measures to baseline network architecture. This data-driven approach advances beyond prior correlational studies by revealing network profiles that may not be captured by linear relationships between connectivity measures and creativity scores. The emergence of these profiles from whole-brain connectivity patterns in healthy individuals, independent of demographic or behavioral information during clustering, suggests the presence of identifiable resting-state network configurations associated with creativity-related characteristics.

### 4.2. Behavioral Correlates of EEG-derived Clusters

Consistent with our second hypothesis, the two EEG-derived groups diverged across multiple dimensions of creativity. Specifically, participants in Cluster 1 reported significantly higher self-perceived creative ability, together with greater engagement in everyday creative activities as measured by the ICAA, suggesting that participants in this cluster not only view themselves as creative but also enact this capacity in daily life. This convergence between subjective efficacy and objective engagement echoes recent multidimensional models of creativity that emphasize the alignment of between self-concept and creative action [62,63].

This group also exhibited higher, albeit non-significant, performances on divergent (DAT) and convergent (Type C and D MAPT) thinking tasks, indicating a consistent directional trend rather than a robust performance advantage. Given that creativity is a multifaceted construct with often dissociable cognitive and behavioral dimensions [64], the dissociation between subjective engagement measures and task performance is particularly noteworthy. Whereas prior research has often identified domain-specific neural correlates of divergent versus convergent thinking, our findings suggest that resting-state network organization may be more closely linked to creative engagement and self-perception than to task-specific performance indices. This interpretation is consistent with the possibility that intrinsic connectivity patterns reflect a general dispositions towards creative engagement, rather than stable cognitive capacity. Replication in larger samples and inclusion of additional task-based EEG measures will be necessary to further clarify these relationships.

Our results also raise an alternative interpretation that was not explicitly anticipated a priori. We initially conceptualized intrinsic functional network architecture as a feature that predisposes individuals toward creativity (i.e., individuals are tuned for creativity). However, an equally plausible account is that sustained and frequent engagement in creative activities may itself shape intrinsic neural connectivity over time, particularly within resting-state alpha dynamics (i.e., individuals are tuned by creativity). From this perspective, the observed connectivity patterns may reflect experience-dependent plasticity associated with creative training and repeated engagement, rather than differences in inherent creative potential. Such engagement may also contribute to elevated creative self-efficacy, as individuals with repeated practice and exposure may develop stronger confidence in their creative abilities. Creative self-efficacy is an affective construct believed to be predictive of creative output and amenable to modulation [65–67], and likely both results from and promotes creative activity engagement and performance. Given the cross-sectional design, these possibilities cannot be disentangled in the present study and remain important avenues for future longitudinal and interventional research.

### 4.3. Reduced Global Connectivity and Alpha-band Hypoconnectivity

Our third hypothesis—that the subgroup scoring higher on creativity measures would show reduced global connectivity—was supported. Participants in Cluster 1 showed broad alpha-band hypoconnectivity relative to their peers. Similar patterns of hypoconnectivity have been reported in individuals with higher creativity scores [26] and creative experts [24], suggesting that lower synchronization at rest is associated with greater cognitive flexibility and associative thinking.

Alpha-band activity, often interpreted as an index of internally oriented attention and cognitive control [68,69], may thus serve as a neural substrate for efficient detachment from external sensory constraints, enabling associative recombination of ideas [70]. From a network efficiency perspective, reduced overall coupling can reflect a less redundant, and potentially more adaptable architecture [71]. In contrast, excessive connectivity, commonly observed in clinical or cognitively rigid populations, has been associated with reduced functional differentiation and impaired reconfiguration capacity [72–74]. Collectively, these observations suggest that the hypoconnectivity observed in Cluster 1 is consistent with a network organisation characterized by adaptive decoupling rather than maximal integration [21], and may serve as a resting-state neural marker associated with creativity-related traits.

### 4.4. Preserved Left-frontal Connectivity and Enhanced Modularity

Against this backdrop of hypoconnectivity, Cluster 1 exhibited relatively preserved left-frontal coupling and greater modular segregation. While sensor-space analyses preclude precise anatomical localization, this pattern suggests that creative individuals maintain stable functional coupling in frontal regions while allowing broader networks to remain functionally differentiated. Frontal regions have long been implicated in cognitive control and constraint relaxation during creative thought [36,75]. Further, left prefrontal engagement facilitates the dynamic switching between generative and evaluative modes during creative thinking [21,76,77].

Higher modularity together with lower local clustering observed in Cluster 1 indicates a network architecture that is globally sparse with preserved modular organization. Modularity reflects the degree to which brain networks can operate semi-independently while maintaining efficient inter-module communication [71,78]. Decreased local clustering implies fewer redundant short-range connections, potentially reducing interference and facilitating long-range integration across modules. This multiscale segregation pattern of reduced local redundancy coupled with enhanced global specialisation has been repeatedly associated with superior cognitive flexibility and creative insight in both EEG [28] and fMRI studies [18,24]. In sum, the observed network organization is consistent with patterns previously linked to flexible cognition, without implying a unique or exclusive configuration of creativity.

### 4.5. Integrative Theoretical Perspectives

The combined pattern of global hypoconnectivity, preserved frontal coupling, and increased modular segregation can be interpreted within existing theoretical frameworks of network efficiency and predictive processing. According to network-neuroscience models [71,79,80], optimal cognitive function arises when the brain operates near a critical balance between integration and segregation, maximising information transfer while maintaining flexibility for rapid reconfiguration [81,82]. Creativity may therefore represent an emergent property of such near-critical dynamics [21,83], where intrinsic connectivity is sufficiently decoupled to enable exploration of novel associations yet sufficiently organised to sustain coherence of thought.

From a predictive-coding perspective [84,85], reduced alpha-band coupling may reflect a resting disposition toward diminished top-down constraint, allowing for greater updating of internal models and recombination of ideas [86,87]. Preserved frontal connectivity could, in turn, anchor this generative process through adaptive regulation of prediction error signals. These interpretations converge with fMRI evidence showing that individuals scoring higher on creativity-related measures exhibit flexible coupling between the default-mode and control networks during ideation [88,89]. Within this context, the present EEG findings provide evidence consistent with large-scale network models of creativity.

### 4.6. Limitations and Future Directions

Several methodological and interpretive limitations should be acknowledged. First, our modest sample size ($N$ = 26), while sufficient to detect large effect sizes in our primary analyses, limits statistical power for detecting more subtle individual differences and constrains generalizability. Nevertheless, the convergence of behavioural and neural clustering provides preliminary support for the robustness of the identified subgroups. Future studies with larger, more diverse samples could use replication-based cross-validation or hierarchical clustering to confirm the stability of these latent profiles.

Second, connectivity was analyzed in sensor space without source localization, which precludes precise anatomical interpretation. While the debiased wPLI approach minimizes volume-conduction-related effects, it does not fully resolve source localisation. Applying source-reconstructed EEG or multimodal fusion with fMRI would help pinpoint cortical and subcortical generators of the observed network configurations and test whether they map onto canonical large-scale brain networks often implicated with creative cognition.

Third, our focus on the alpha-band, while theoretically motivated and methodologically sound, represents only one aspect of the brain's oscillatory repertoire. Creativity likely involves interactions across multiple frequency bands [90,91], and most importantly, the brain oscillations are unlikely to be independent during creative ideation [92,93]. Future work should apply innovative methods to examine whether similar organizational principles extend to other frequency bands, particularly theta (implicated in memory and associative processing) and gamma (linked to feature binding and conscious processing), and cross-frequency coupling.

Fourth, the resting-state condition may itself be heterogenous across individuals. Alpha power and connectivity are known to vary with fluctuations in vigilance, mind-wandering and internally directed attention. Although the behavioral dissociation between clusters argues against a purely state-dependent account, integrating pupillometry, subjective experience sampling, or concurrent vigilance metrics would allow state-trait disambiguation.

Finally, while we interpret hypoconnectivity as reflecting neural efficiency and flexibility, alternative explanations are plausible. Reduced coupling could indicate greater neural independence but also weaker communication efficiency or idiosyncratic network organisation. Longitudinal and intervention studies, such as creativity training or neurofeedback paradigms, could test whether intrinsic connectivity patterns are malleable and whether changes in modularity or coupling predict subsequent creative development.

## 5. CONCLUSION

This study provides convergent evidence that individual differences in creativity-related characteristics are associated with large-scale topology of resting-state EEG networks. Using an unsupervised, data-driven approach, we identified two participant groups defined by distinct patterns of alpha-band connectivity that were associated with systematic differences across multiple dimensions of creativity, including creative activity engagement and creative self-efficacy, alongside consistent but non-significant trends for divergent and convergent thinking performance.

The combination of broad alpha-band hypoconnectivity, relatively preserved left-frontal coupling, and enhanced modular segregation in the higher-scoring subgroup is consistent with a network organization characterized by reduced redundancy and increased functional differentiation. Such an organisation may maintain sufficient functional decoupling to enable exploratory associative processing while preserving the structural coherence necessary for controlled evaluation.

These findings extend network-based accounts of creativity by providing temporally resolved EEG evidence that complements prior fMRI work indicating that creativity is expressed not only through the anatomical connectivity of large-scale brain systems but also through the rapid oscillatory dynamics that coordinate their activity. More broadly, this work illustrates the potential of unsupervised machine-learning frameworks applied to resting-state neurophysiology for identifying reproducible patterns of neurocognitive variation.

Within the context of emerging AI applications in cognitive neuroscience and precision medicine, this data-driven framework offers a scalable and non-invasive approach to characterizing individual differences in brain organisation. Future studies incorporating larger samples, source-space reconstruction, dynamic-connectivity metrics, and multimodal neuroimaging will be crucial for validation and extension. Taken together, the present results indicate that resting state EEG network organization captures meaningful variation associated with creative engagement and self-efficacy, even in the absence of explicit task demands.

## Data Availability

The data that support the findings of this study are available on request from the corresponding author. The data are not publicly available due to privacy or ethical restrictions.

**Author Contributions**

**Samir Damji:** conceptualization, investigation, data curation, formal analysis, visualization, writing – original draft, funding acquisition. **Simrut Kurry:** investigation, data curation. **Shazia'Ayn Babul:** conceptualization, methodology. **Joydeep Bhattacharya:** conceptualization, supervision, writing – review & editing. **Naznin Virji-Babul:** conceptualization, supervision, project administration, funding acquisition, writing – review & editing.

## Figure legends

**Figure 1.** *Analysis pipeline for clustering resting-state EEG based on graph similarity.*

(A) 64 x 64 functional connectivity (FC) matrices generated across subjects using debiased wPLI.
(B) Pairwise graph distances computed for weighted, undirected networks using the DeltaCon algorithm.
(C) Dimensionality reduction performed with UMAP to obtain a 2D embedding.
(D) K-means clustering applied, and the optimal cluster solution determined using the silhouette coefficient.
(E) Behavioral and rsEEG features statistically compared across clusters.

**Figure 2.** *Graph similarity analysis results.*

(A) Pairwise DeltaCon distance matrix, where blue values indicate more similar graph structures and red values indicate more dissimilar structures, as indexed by fast belief propagation.
(B) Two-dimensional UMAP embedding of DeltaCon distances for visualization, followed by k-means clustering on the reduced coordinates.
(C) The silhouette method identified $k = 2$ as the optimal cluster solution (candidate values restricted to $k \leq 5$ due to sample size).
(D) DeltaCon distance matrix with rows and columns reordered by cluster assignment. Blue colors along the diagonal reflect within-cluster similarity, while warmer colors off the diagonal reflect between-cluster dissimilarity.
(E) Clustering robustness assessed using Adjusted Rand Index (ARI) across random seeds. Violin plot distributions are shown for 100 perturbations of the full UMAP–k-means pipeline (left) and for k-means clustering with fixed UMAP embeddings (right), indicating high algorithmic clustering stability and moderate pipeline-level variability attributable to embedding stochasticity. Solid black lines denote mean ARI values; dotted red lines denote median ARI values; dashed grey lines indicate chance-level agreement.

**Figure 3.** *Behavioral creativity profiles across the two clusters.*

Distinct behavioral creativity profiles emerged across Cluster 1 (red) and Cluster 0 (blue), groups derived from resting-state EEG network structure. Medium-to-large effect sizes were observed across all measures, with significant differences observed for SR of creative ability and the ICAA Activities subscale.

**Figure 4.** *Mean functional connectivity difference matrix (Cluster 1 – Cluster 0).*

Warmer values (red) indicate stronger connectivity in Cluster 1, while cooler values (blue) indicate stronger connectivity in the Cluster 0. The FC difference matrix (64 x 64 channels) reveals broad hypoconnectivity in Cluster 1, consistent with significantly reduced mean wPLI compared to Cluster 0.

**Figure 5.** *Topographical edgewise and nodal connectivity differences between clusters.*

(A) Statistically significant edgewise FC differences after FDR correction ($\alpha = 0.03$). Blue edges indicate stronger connectivity in Cluster 0, while red edges (none observed) would have indicated stronger connectivity in Cluster 1. Line thickness is proportional to the magnitude of Cohen's d, highlighting that all ten surviving edges showed large effect sizes. The predominance of blue edges provides edge-level evidence for widespread hypoconnectivity in Cluster 1.

(B) Node strength difference map (Cluster 1 – Cluster 0). All values were negative, indicating reduced node connectivity in Cluster 1 relative to Cluster 0. Warmer colors represent relatively preserved connectivity (smaller decreases in node strength) in Cluster 1, most notably over the left anterior region, while cooler colors reflect greater reductions. This node-level perspective reinforces the presence of distributed reductions in resting-state synchronization in individuals within Cluster 1.

**Figure 6.** *Distributions of global graph metrics within EEG clusters.*

Histograms depict the distribution of five weighted global graph metrics across participants in Cluster 1 (red) and Cluster 0 (blue). Participants in Cluster 1 showed reduced node strength, global efficiency, and clustering coefficient, and increased modularity and characteristic path length. All group differences reached statistical significance based on two-tailed Welch's *t*-tests. Asterisks denote significance levels ($p < .05$, $p < .01$, $p < .001$). Effect sizes (Cohen's *d*) were large across all metrics (≈1.0–2.2).